\documentclass[10pt]{article}

\usepackage{graphicx,amsmath,amsfonts,amssymb,dcolumn,amsthm,euscript,cite}

\usepackage{esint}
\usepackage{lscape}
\usepackage{pstricks}
\usepackage{color}
\usepackage{braket,euscript}
\usepackage{fancyvrb}

\renewcommand{\d}{\ensuremath{\mathrm{d}}}

\renewcommand{\d}{\ensuremath{\mathrm{d}}}

\newcommand{\dd}{{\textrm{d}}}

\newcommand{\bea}{\begin{eqnarray}}	
\newcommand{\eea}{\end{eqnarray}}
\newcommand{\be}{\begin{equation}}	
\newcommand{\ee}{\end{equation}}
\newcommand{\beq}{\begin{equation}}	
\newcommand{\eeq}{\end{equation}}

\begin{document}

\title{\textbf{ The universal character of Zwanziger's horizon function in Euclidean Yang--Mills theories}}

\author{\textbf{M.~A.~L.~Capri$^1$}\thanks{caprimarcio@gmail.com},
\textbf{D.~Dudal$^{2,3}$}\thanks{david.dudal@kuleuven.be},
\textbf{M.~S.~Guimaraes$^1$}\thanks{msguimaraes@uerj.br}, \\
\textbf{A.~D.~Pereira$^4$}\thanks{a.pereira@thphys.uni-heidelberg.de},
\textbf{B.~W.~Mintz$^1$}\thanks{bruno.mintz@uerj.br},
\textbf{L.~F.~Palhares$^1$}\thanks{leticia.palhares@uerj.br},
\textbf{S.~P.~Sorella$^1$}\thanks{silvio.sorella@gmail.com},\\\\\
\textit{{\small $^1$UERJ -- Universidade do Estado do Rio de Janeiro,}}\\
\textit{{\small Instituto de F\'isica -- Departamento de F\'{\i}sica Te\'orica -- Rua S\~ao Francisco Xavier 524,}}\\
\textit{{\small 20550-013, Maracan\~a, Rio de Janeiro, Brasil}} \\\
\textit{{\small $^2$ KU Leuven Campus Kortrijk -- Kulak, Department of Physics, }} \\
\textit{{\small Etienne Sabbelaan 53 bus 7657, 8500 Kortrijk, Belgium}}\\
\textit{{\small $^3$ Ghent University, Department of Physics and Astronomy, }}\\
\textit{{\small Krijgslaan 281-S9, 9000 Gent, Belgium}} \\\
\textit{{\small $^4$ Institute for Theoretical Physics, University of Heidelberg,}}\\
\textit{{\small Philosophenweg 12, 69120 Heidelberg, Germany}}}
\date{}

\maketitle

\begin{abstract}
In  light  of the  recently established  BRST invariant formulation of the Gribov--Zwanziger theory, we show that Zwanziger's horizon function  displays a universal character. More precisely, the correlation functions of local BRST invariant operators evaluated with the Yang--Mills action supplemented with  a BRST invariant version of the Zwanziger's horizon function and quantized in an arbitrary class of covariant, color invariant and  renormalizable gauges which reduce to the Landau gauge when all gauge parameters are set to zero, have a unique, gauge parameters independent result, corresponding to that of the Landau gauge  when the restriction to the Gribov region $\Omega$ in the latter gauge is imposed. As such, thanks to the BRST invariance, the cut-off at the Gribov region $\Omega$ acquires a gauge independent meaning in the class of the physical correlators.
\end{abstract}


\newpage



\section{Introduction: a short review of Zwanziger's horizon function}
The so-called Gribov--Zwanziger theory \cite{Vandersickel:2012tz}  handles the issue of the Gribov copies \cite{Gribov:1977wm}  in the Landau gauge, $\partial_\mu A^a_\mu=0$, by restricting the domain of integration in the functional integral to the Gribov region $\Omega$
\begin{equation}
\Omega = \left\{\,A^a_\mu\,, \,\,\partial_\mu A^a_\mu = 0\,, \; {\cal M}^{ab}(A)= -\partial_\mu D^{ab}_\mu(A) >0\right\}\,,
\label{omega}
\end{equation}
where ${\cal M}^{ab}(A)$ is the Faddeev-Popov operator\footnote{Due to the Landau condition, $\partial_\mu A^a_\mu=0$, the Faddeev-Popov operator ${\cal M}^{ab}(A)=-\partial_\mu D^{ab}_\mu(A)$  is  Hermitian, a crucial property for the definition of $\Omega$.}  and $D^{ab}_\mu = \delta^{ab}\partial_\mu - gf^{abc}A^{c}_{\mu}$ stands for the covariant derivative in the adjoint representation of $SU(N)$.   For the partition function in  $d=4$  Euclidean space, one writes

\begin{equation}
{\cal Z} = \int_{\Omega} \left[\EuScript{D}A \right] \; \delta(\partial A) \; (\det {\cal M})\;  \mathrm{e}^{-S_{\text{YM}}}\,, \qquad  S_{\text{YM}}= \frac{1}{4} \int \dd^4x\; F^{a}_{\mu\nu}  F^{a}_{\mu\nu}  \;, \label{z1}
\end{equation}
The restriction to the region $\Omega$ has been put on firm basis due to the following properties \cite{Dell'Antonio:1991xt}: \textit{i)} $\Omega$ is bounded in all directions in field space. The boundary, $\partial \Omega$,  of $\Omega$, where the first vanishing eigenvalue of the Faddeev-Popov operator  ${\cal M}^{ab}(A)$ shows up, is the first Gribov horizon; \textit{ii)} $\Omega$  is convex; \textit{iii)} All gauge orbits cross $\Omega$ at least once. In particular, the latter implies that gauge configurations lying outside the region $\Omega$ are copies of configurations belonging to $\Omega$, giving thus a well  motivated  support to expression \eqref{z1},  in the sense that it does take into account all physically different gauge configurations \footnote{It must be pointed out that the region $\Omega$ is not  free from Gribov ambiguities. Additional Gribov copies still exist inside $\Omega$  \cite{semenov,vanBaal:1991zw}.  A smaller region contained within $\Omega$,  known as the  fundamental modular region, does exist which is fully free from Gribov copies \cite{vanBaal:1991zw}.  Ideally, one should restrict the domain of integration in the functional integral to the fundamental modular region rather than to the Gribov region. Nevertheless, until now,  a practical  way to restrict the path integral to  the fundamental modular region is not yet at our disposal. We focus thus on the  Gribov region  $\Omega$.}.  \\\\  At the practical computational level,  Zwanziger's horizon function $H(A)$ \cite{Vandersickel:2012tz,Zwanziger:1989mf}  plays a pivotal role in the evaluation of the partition function \eqref{z1}. It turns out that, in eq.\eqref{z1},  modulo some assumptions and simplifications, the restriction to the region $\Omega$ can be lifted inside the Boltzmann weight through the addition of a novel term, given precisely by $H(A)$. In other words, expression \eqref{z1} can be rewritten as
\begin{equation}
{\cal Z} =  \int \left[\EuScript{D}A\right]  \; \delta(\partial A) \; (\det {\cal M})\;
\mathrm{e}^{-(S_{\text{YM}}
+\gamma^4H(A)-4V\gamma^4(N^2-1))}\,,
\label{z2}
\end{equation}
where
\begin{equation}
H(A)=g^2\int \dd^4x \dd^4y~f^{abc}A^{b}_{\mu}(x)\left[{\cal M}^{-1}\right]^{ad}(x,y)f^{dec}A^{e}_{\mu}(y)\,,
\label{intro8}
\end{equation}
is the horizon function, $V$ is the space-time volume and $N$ is the number of colors. The parameter $\gamma$ has mass dimension one and is called the Gribov parameter. It is not free, being determined through a gap equation, known as the horizon condition:
\begin{equation}
\langle H(A)\rangle=4V(N^2-1)\,,
\label{intro9}
\end{equation}
where the expectation value $\langle H(A)\rangle$ has to be evaluated with the measure defined by expression \eqref{z2}. \\\\It is worth mentioning here that in his seminal paper \cite{Gribov:1977wm}, Gribov implemented the restriction to $\Omega$ only at the first order, by following a different path than that outlined by Zwanziger in \cite{Vandersickel:2012tz,Zwanziger:1989mf}. More precisely, in \cite{Gribov:1977wm}, the restriction to $\Omega$  was worked out by means of the so-called Gribov no-pole condition, amounting to  require that the inverse,  ${{\cal M}^{-1}}$, of the Faddeev-Popov operator  is strictly positive at zero momentum,  under the assumption that the lowest value of it is achieved at zero momentum. At first look, Gribov's no-pole condition and Zwanziger's horizon condition might appear different. Nevertheless, it turns out that Gribov's no-pole condition can be resummed to all order \cite{Gomez:2009tj,Capri:2012wx}, so that a closed expression can be found. The output of the resummation yields exactly  Zwanziger's horizon function, eq.\eqref{intro8},  and  the horizon condition, eq.\eqref{intro9},  indicating thus the equivalence  of both methods.\\\\ As it is apparent  from eq.\eqref{intro8}, the horizon function $H(A)$ is non-local, giving rise to  a non-local action. However,  expression \eqref{z2} can be cast in  local form \cite{Vandersickel:2012tz,Zwanziger:1989mf}  through the introduction of a pair of bosonic auxiliary fields  $(\bar{\varphi},\varphi)^{ab}_\mu$ and a pair of anticommuting fields $(\bar{\omega},\omega)^{ab}_\mu$. The resulting local action is known as the Gribov--Zwanziger (GZ) action $S_{\text{GZ}}$, being given by
\begin{equation}
S_{\text{GZ}} = S_{\text{YM}}+S_{\text{FP}}^{\text{Landau}}-\int \dd^4x\left(\bar{\varphi}^{ac}_{\mu}{\cal M}^{ab}(A) {\varphi}^{bc}_{\mu}-\bar{\omega}^{ac}_\mu{\cal M}^{ab}(A) \omega^{bc}_\mu
+\gamma^{2} ~gf^{abc}A^{a}_{\mu}(\varphi+\bar{\varphi})^{bc}_\mu \right) \,,
\label{intro10}
\end{equation}
where $S_{\text{FP}}^{\text{Landau}}$ stands for the Faddeev-Popov action of the Landau gauge, {\it i.e.}
\begin{equation}
S_{\text{FP}}^{\text{Landau}}=\int \dd^4x\left(ib^a\partial_{\mu}A^a_\mu + \bar{c}^a\partial_\mu D^{ab}_\mu c^b\right)\,.   \label{fpl}
\end{equation}
Therefore, for the partition function, eqs.\eqref{z1},\eqref{z2}, we have
\begin{equation}
 {\cal Z} =  \int \left[\mathcal{D}\Phi\right]  \;
\mathrm{e}^{-(S_{\text{GZ}} -4V\gamma^4(N^2-1))}=\mathrm{e}^{-V  {\cal E}_{v}}\,,
\label{z3}
\end{equation}
where $\left[\mathcal{D}\Phi\right] $ is a short-hand notation for integration over all fields appearing in the Boltzmann weight of eq.\eqref{z3}, namely: $A_{\mu},b, c,\bar c,\varphi,\bar{\varphi},\omega,\bar{\omega}$. In the local formulation, the horizon condition \eqref{intro9} defining the Gribov parameter $\gamma$ reads $\left.\frac{\partial {\cal E}_{v}}{\partial \gamma^2}\right|_{\gamma^2\neq0}=0$. The Gribov--Zwanziger action, eq.\eqref{intro10}, turns out to be  multiplicatively  renormalizable to all orders  \cite{Vandersickel:2012tz,Zwanziger:1989mf,Maggiore:1993wq}, implying that explicit calculations can be carried out in a consistent way. \\\\In particular, for the correlation functions of local gauge invariant operators  ${\cal O}(x)$, we may write
\begin{equation}
\langle {\cal O}(x) {\cal O}(y) \rangle\Big|_{S_{\text{GZ}}}^{\text{Landau}}  =  \frac{ \int \left[\mathcal{D}\Phi\right] \; {\cal O}(x) {\cal O}(y)  \;
\mathrm{e}^{-(S_{\text{GZ}} -4V\gamma^4(N^2-1))} }{\int \left[\mathcal{D}\Phi\right]  \;
\mathrm{e}^{-(S_{\text{GZ}} -4V\gamma^4(N^2-1))} }   \;, \label{oo}
\end{equation}
where  ${\cal O}(x)$ \footnote{From the general results on the cohomology of the BRST operator in Yang--Mills theories, it follows that the set $\{{\cal O}(x)\}$ is spanned by local colorless operators of arbitrary  dimensions built up with the field strength $F^{a}_{\mu\nu}$ and its covariant derivative $D^{ab}_\mu$, see \cite{Piguet:1995er} and refs.~ therein. Fermions can also be added when necessary.} stands for a generic  gauge invariant operator. Correlation functions of this type are of fundamental importance in order to unravel the physical content of the restriction to the Gribov region and of the Gribov--Zwanzgier action. For instance, expression \eqref{oo} can be directly employed to study the spectrum of the theory, as done in the case of the glueballs in \cite{Baulieu:2009ha,Dudal:2010cd,Dudal:2013wja}. \\\\Nevertheless, a drawback of the original Gribov--Zwanziger framework, eqs.\eqref{intro10},\eqref{z3}, is the lack of BRST invariance \cite{Vandersickel:2012tz}. As it stands, the action \eqref{intro10} exhibits a soft breaking of the BRST invariance, which turns out to be  proportional to the Gribov parameter $\gamma$ \cite{Vandersickel:2012tz}. Although this feature does not jeopardize  the renormalizability of the Gribov--Zwanziger action \eqref{intro10}, it obscures the physical meaning of $\gamma$ itself, which encodes the restriction to the region $\Omega$. Furthermore, the lack of BRST invariance does not make  evident  a natural  extension of the Gribov--Zwanziger  setup to other covariant renormalizable gauges as, for example, the linear covariant gauges. Finally, without BRST invariance one is not able to prove that the correlation functions of gauge invariant operators are independent of the gauge parameters entering the gauge fixing condition, a fundamental property in order to attach a physical meaning to expression \eqref{oo}. Needless to say, the issue of the BRST symmetry and of its soft breaking in the Gribov--Zwanziger theory has been  object of intensive investigations,   see  \cite{Maggiore:1993wq,Serreau:2012cg,Serreau:2015yna,Capri:2014bsa,Dudal:2009xh,Sorella:2009vt,Baulieu:2008fy,Capri:2010hb,Dudal:2012sb,Dudal:2014rxa,Pereira:2013aza,Pereira:2014apa,Lavrov:2011wb,Lavrov:2013boa,Moshin:2015gsa,Schaden:2014bea,Cucchieri:2014via}. \\\\Recently, we have been able to reformulate the Gribov--Zwanziger theory in such a way that a manifest exact BRST invariance could be established \cite{Capri:2015ixa,Capri:2016aqq,Capri:2016gut,Capri:2017bfd,Dudal:2017jfw}.  The details of this construction will be shortly reviewed in the next section. The existence of an exact BRST symmetry  provides a clear physical meaning to the Gribov parameter $\gamma$, while allowing us to establish that the correlation functions  of gauge invariant operators are independent of the gauge parameters. These features give a universal, gauge independent, character to Zwanziger's horizon function $H(A)$, eq.\eqref{intro8},  albeit after replacing it with its BRST invariant counterpart, see the next section. \\\\Let us also mention that, as observed in \cite{Dudal:2007cw,Dudal:2008sp,Dudal:2011gd,Gracey:2010cg},
the restriction to the Gribov $\Omega$ region leads to additional non-perturbative  instabilities giving rise to the formation of dimension-two condensates, namely  $\langle A^{a}_{\mu} A^{a}_{\mu}\rangle$, whose value, in the presence of Gribov's horizon, supplements the one also present in perturbative YM \cite{Verschelde:2001ia}, and $\langle\bar{\varphi}^{ab}_{\mu}\varphi^{ab}_{\mu}-\bar{\omega}^{ab}_{\mu} \omega^{ab}_{\mu}\rangle$. Taking into account the existence of such condensates from the beginning, gives rise to the so-called Refined Gribov--Zwanziger (RGZ) action, given by
\begin{equation}
S_{\text{RGZ}}=S_{\text{GZ}}+\frac{m^2}{2}\int \dd^4x~A^{a}_{\mu}A^{a}_{\mu}-M^2\int \d^4x\left(\bar{\varphi}^{ab}_{\mu}\varphi^{ab}_{\mu}-\bar{\omega}^{ab}_{\mu}\omega^{ab}_{\mu}\right)\,,
\label{intro11}
\end{equation}
where, as much as the Gribov parameter $\gamma$,  the mass parameters $m$ and $M$ are  not  free, but dynamically  determined by minimizing their respective effective action, see \cite{Dudal:2011gd}. As the GZ action, the RGZ action is renormalizable to all orders in perturbation theory \cite{Dudal:2008sp}. In particular, the tree-level gluon propagator stemming from \eqref{intro11} attains a finite value at $k=0$. Such a  behavior is in agreement with ruling  lattice data as well as with functional and effective methods, see \cite{Cucchieri:2007rg,Oliveira:2012eh,Duarte:2016iko,Aguilar:2008xm,Fischer:2008uz,Boucaud:2011ug,Tissier:2010ts,Weber:2011nw,Cyrol:2016tym,Reinosa:2017qtf,Siringo:2015wtx,Frasca:2007uz,Chaichian:2018cyv,Gao:2017uox,Huber:2015ria,Aguilar:2015nqa,Aguilar:2016ock,Maas:2017csm} for a non-exhaustive list.  \\\\Let us end this short summary on the Gribov--Zwanziger formulation by outlining the organization of the paper. Sect.2 is devoted to the  BRST invariant reformulation of both GZ and RGZ actions and to its consequences on the Gribov parameter $\gamma$ as well as on the correlation functions \eqref{oo}. In Sect.3 we  present the main results of this paper: a  generalization of the Gribov--Zwanziger setup to an arbitrary class of covariant, color invariant and renormalizable  gauge fixings which reduce to the Landau gauge when setting the gauge parameters to zero, providing thus a universal character to (the BRST invariant extension of) Zwanziger's horizon function. Sect.4 collects our conclusion.

\section{BRST invariant reformulation of the GZ theory }

The main tool of the BRST invariant reformulation \cite{Capri:2015ixa,Capri:2016aqq,Capri:2016gut,Capri:2017bfd} of the Gribov--Zwanziger theory has been the use of a gauge invariant and transverse field configuration $A^h_\mu$, obtained
 by minimizing the functional
$\mathrm{Tr}\int d^{4}x\,A_{\mu }^{u}A_{\mu }^{u}$ along the gauge
orbit of $A_{\mu }$ \cite{Dell'Antonio:1991xt,Zwanziger:1990tn}, namely
\begin{equation}
A_{\min }^{2} \equiv \min_{\{u\}}\mathrm{Tr}\int d^{4}x\,A_{\mu
}^{u}A_{\mu }^{u}\;,\qquad\mathrm{with}\qquad
A_{\mu }^{u} = u^{\dagger }A_{\mu }u+\frac{i}{g}u^{\dagger }\partial _{\mu
}u\;.    \label{Aminn0}
\end{equation}
In particular, looking at  the stationary condition of the functional \eqref{Aminn0},  one gets a non-local transverse
field configuration $A^h_\mu$, $\partial_\mu A^h_\mu=0$, which can be expressed as an infinite series in
the gauge field $A_\mu$, see Appendix~A of \cite{Capri:2015ixa}, {\it i.e.}
\begin{eqnarray}
A_{\mu }^{h} &=&\left( \delta _{\mu \nu }-\frac{\partial _{\mu }\partial
_{\nu }}{\partial ^{2}}\right) \phi _{\nu }\;,  \qquad  \partial_\mu A^h_\mu= 0 \;, \nonumber \\
\phi _{\nu } &=&A_{\nu }-ig\left[ \frac{1}{\partial ^{2}}\partial A,A_{\nu
}\right] +\frac{ig}{2}\left[ \frac{1}{\partial ^{2}}\partial A,\partial
_{\nu }\frac{1}{\partial ^{2}}\partial A\right] +O(A^{3})\;.  \label{min0}
\end{eqnarray}
Remarkably,  the configuration $A_{\mu }^{h}$ turns out to be left invariant
by infinitesimal gauge transformations order by order in the gauge coupling $g$, see  \cite{Lavelle:1995ty,Capri:2015ixa}:
\begin{equation}
\delta A_{\mu }^{h} = 0\;,  \qquad\mathrm{with}\qquad
\delta A_{\mu } = -\partial _{\mu }\omega +ig\left[ A_{\mu },\omega \right]
\;.  \label{gio}
\end{equation}
The infinite series \eqref{min0} is an expansion in powers of the coupling constant $g$. As such, its meaning is that of a weak coupling expansion, as it will be stated in more precise terms below.  \\\\Moreover, as one directly observes from eq.\eqref{min0}, a divergence $(\partial_\mu A_\mu)$ is always present in  all higher order
terms  \cite{Lavelle:1995ty,Capri:2015ixa}.  Therefore, we can rewrite Zwanziger's horizon function $H(A)$ in terms of the invariant field $A^h_\mu$  \cite{Capri:2015ixa}, namely\footnote{It is important to emphasize that we do not perform a variable transformation between $A$ and $A^h$. In fact, as written in eq.\eqref{min0}, $A^ h$ is a function of the field $A$. In eq.\eqref{hr} we add all the (highly non-local) structure to $H(A)$ to compose $H(A^h)$ and subtract it as indicated by the term $R(A)\partial A$. The nontrivial feature is that it is possible to extract a factor $\partial A$ as explained in \cite{Capri:2015ixa}. Hence, since there is no change of variables implemented so far, eq.\eqref{hr} does not entail any Jacobian in the functional integral.}
\begin{equation}
H(A) = H(A^h)- R(A) (\partial A)    \;, \label{hr}
\end{equation}
where $R(A)(\partial A)$  is a short-hand notation for $R(A)(\partial A) = \int d^4x d^4x R^a(x,y) (\partial_\mu A^a_\mu)(y)$, with $R(A)$ being an infinite non-local power series in $A_\mu$, and
\begin{equation}
H(A^h)=g^2\int \dd^4x \dd^4y~f^{abc}A^{h,b}_{\mu}(x)\left[{\cal M}^{-1}(A^h)\right]^{ad}(x,y)f^{dec}A^{h,e}_{\mu}(y)\,.
\label{Hh}
\end{equation}
Furthermore, following   \cite{Capri:2015ixa}, the term $R(A)(\partial A)$ can be fully reabsorbed through a redefinition of the Lagrange multiplier $b$, {\it i.e.}
\begin{equation}
b \rightarrow b + i \gamma^4 R(A) \;, \label{brd}
\end{equation}
which has unity Jacobian. Thus, expression  \eqref{intro10}   becomes \cite{Capri:2015ixa}
\begin{equation}
S_{\text{GZ}} = S_{\text{YM}}+S_{\text{FP}}^{\text{Landau}}-\int \dd^4x\left(\bar{\varphi}^{ac}_{\mu}{\cal M}^{ab}(A^h) {\varphi}^{bc}_{\mu}-\bar{\omega}^{ac}_\mu{\cal M}^{ab}(A^h) \omega^{bc}_\mu
+\gamma^{2} ~gf^{abc}A^{h,a}_{\mu}(\varphi+\bar{\varphi})^{bc}_\mu \right) \,,
\label{sgzn}
\end{equation}
and the corresponding partition function remains with exactly the same measure as before, in e.g. eq.\eqref{z3}.\\\\ We are now left with the issue of localizing the operator $A^h_\mu$, a task which has been successfully handled in \cite{Capri:2016aqq}, yielding  the local expression
\begin{eqnarray}
S_{\text{GZ}}^{\text{loc}} &=&S_{\text{YM}}+S_{\text{FP}}^{\text{Landau}}-\int \d^4x\left(\bar{\varphi}^{ac}_{\mu}{{\cal M}^{ab}}(A^h) {\varphi}^{bc}_{\mu}-\bar{\omega}^{ac} {\cal M}^{ab}(A^h)\omega^{bc}_\mu\right) \nonumber \\
\; \; \; \; &-&\gamma^{2}\int \d^4x~gf^{abc}(A^h)^a_{\mu}(\varphi+\bar{\varphi})^{bc}_\mu
+\int \d^4x~\left( \tau^a \partial_\mu (A^h)^a_\mu - \bar{\eta}^a {\cal M}^{ab}(A^h)\eta^b \right) \,,
\label{sgzl}
\end{eqnarray}
where
\begin{equation}
{\cal M}^{ab}(A^h) =    -\partial_\mu D^{ab}_\mu(A^h) \,,
\label{mh}
\end{equation}
and
\begin{equation}
A^{h}_{\mu} = h^\dagger A_\mu h +\frac{i}{g}h^\dagger \partial_\mu h\,, \qquad  h = \mathrm{e}^{ig\xi^a T^a}\,,
\label{intro17}
\end{equation}
with $\xi$ being an  auxiliary  localizing Stueckelberg field and $T^a$ are the generators of $SU(N)$. In this local version, the partition function of the theory now entails a new functional measure, including not only the measure from eq.\eqref{z3} but also the integration over the new localizing fields $\xi,\tau,\eta,\bar{\eta}$.\\\\According to eq.\eqref{min0},  equation \eqref{intro17} has to be understood as a powers series in $\xi^a$, namely
\begin{equation}
A^{h,a}_\mu  = A^a_\mu - D^{ab}_\mu \xi^b  - \frac{g}{2} f^{abc} \xi^b D^{cd}_\mu \xi^d + O(\xi^3)  \;, \label{exA}
\end{equation}
meaning that the whole action \eqref{sgzl} contains an infinite series of terms in  powers  of $\xi$. These terms can be seen as parameterizing weak coupling fluctuations above a nontrivial non-perturbative vacuum, encoded in the Gribov parameter $\gamma^2$. We stress here that, although the expression \eqref{exA} is non-polynomial in the sense that it is an infinite power series, all its terms are local products of fields with only one derivative\footnote{Note that the single covariant derivative $D_\mu^{ab}$ already saturates the dimension of $A^{h,a}_\mu $, so that it can only appear linearly in \eqref{exA}, even for terms of higher order in $\xi$. This is also evident from the gauge transformation law \eqref{intro17} which counts one derivative. A generic gauge transformation is also a local but non-polynomial expression. } and the resulting action involves therefore terms up to second order in derivatives. Thus, the action containing $A^{h,a}_\mu$  as written in \eqref{exA} is effectively local, and as such the usual theorems of local quantum field theory apply to it. \\\\That the action \eqref{sgzl} gives a local setup for the non-local operator $A^h_\mu$ of eq.\eqref{min0} follows by noticing that upon using the equation of motion of the Lagrange multiplier $\tau$, i.e.~the transversality constraint
 \begin{equation}
 \partial_\mu A^h_\mu = 0  \;, \label{tr}
 \end{equation}
we can solve iteratively for the Stueckelberg field, see Appendix~A of \cite{Capri:2015ixa}:
\begin{equation}
\xi_s = \frac{1}{\partial^2} \partial A + \frac{ig}{\partial^2} [\partial A, \frac{\partial A}{\partial^2}] + \cdots \;, \label{xis}
\end{equation}
Inserting  eq.\eqref{xis}  in eq.~\eqref{exA} yields  back the non-local version \eqref{min0}. The extra ghosts $({\bar \eta}, \eta)$ account for the Jacobian arising from the functional integration over $\tau$ which gives a delta-function of the type  $\delta(\partial A^h)$. \\\\The local action $S_{\text{GZ}}^{\text{loc}}$, eq.\eqref{sgzl},  enjoys an exact nilpotent BRST symmetry, see \textit{e.g.} \cite{Capri:2017bfd}, thanks to the BRST-invariance of $A^h$, \textit{i.e.} $sA^h =0$, with $s$ the BRST operator, see \cite{Capri:2015ixa,Capri:2016aqq,Capri:2016gut,Capri:2017bfd} for details and for the generalization to the Refined Gribov-Zwanziger case. Despite the use of a dimensionless localizing Stueckelberg field $\xi^a$, both actions \eqref{sgzl} and its refined version have been proven to be renormalizable to all orders, thanks to the pivotal role played by the transversality constraint \eqref{tr} and to the powerful Slavnov-Taylor identities following from the BRST invariance, see \cite{Capri:2017bfd,Fiorentini:2016rwx,Capri:2017npq} for a detailed proof by means of the algebraic renormalization  framework \cite{Piguet:1995er}. \\\\We are now ready to exploit a few properties of the BRST invariant reformulation of the Gribov--Zwanziger framework. The first important consequence is that the Gribov parameter $\gamma$ can be given a clear physical meaning, being a nontrivial BRST invariant parameter, as expressed by
\begin{equation}
s\left( \frac{\partial S_{\text{GZ}}^{\text{loc}}}{\partial \gamma^2} \right)= - s  \int \d^4x~gf^{abc}(A^h)^a_{\mu}(\varphi+\bar{\varphi})^{bc}_\mu = 0 \;, \qquad s\left( \frac{\partial S_{\text{GZ}}^{\text{loc}}}{\partial \gamma^2} \right) \neq s {\hat \Delta} \;, \label{coh1}
\end{equation}
for any local polynomial ${\hat \Delta}$. Equations  \eqref{coh1} state that the Gribov parameter $\gamma^2$ is associated with a nontrivial element of the cohomology of the BRST operator $s$, just as the coupling constant $g^2$ or the  bare quark masses if present.  As such, it has the meaning of a physical parameter  that can enter  the correlation functions of the local gauge invariant operators. In fact, as it will be shown in the next section, eqs.\eqref{coh1} imply that $\gamma^2$ is independent of the gauge parameters entering the gauge fixing condition, a necessary feature in order to have the meaning of a physical quantity.  Of course, equations similar to \eqref{coh1}  hold for the parameters $(m^2,M^2)$ entering the RGZ action.  \\\\A second important property which can be derived is the equivalence of the physical correlation functions evaluated with the original Gribov--Zwanziger action, eq.\eqref{intro10}, and with its local BRST invariant formulation, eq.\eqref{sgzl}, namely
\begin{equation}
\langle {\cal O}(x) {\cal O}(y) \rangle\Big|_{S_{\text{GZ}}}^{\text{Landau}} = \langle {\cal O}(x) {\cal O}(y) \rangle\Big|_{{S_{\text{GZ}}^{\text{loc}}}}^{\text{Landau}}  \;, \label{equiv}
\end{equation}
where $S_{\text{GZ}}$ and ${S_{\text{GZ}}^{\text{loc}}}$ are given by eqs.\eqref{intro10} and \eqref{sgzl}, respectively.  This is an important consequence of the reformulation of the GZ action. It tells us that the result obtained within the novel local BRST invariant formulation is precisely the same as that obtained with the original GZ action\footnote{Although in eq.~\eqref{equiv} gauge invariant quantities ${\cal O}(x)$ are considered, we underline that the equivalence \eqref{equiv} holds for more general correlation functions as, for example, the $n$-point gluon correlators $\langle A_{\mu_1}(x_1) \ldots A_{\mu_n}(x_n)\rangle$.  It holds in fact for all correlators excluding those explicitly containing the $b$-field.}. Said otherwise, the novel formulation is completely equivalent to the original one.  Moreover, eq.\eqref{equiv} will play a key role in the forthcoming discussion on the universality character of Zwanziger horizon function  $H(A^h)$, eq.\eqref{Hh}. \\\\Due to its relevance, let us give a detailed look at eq.\eqref{equiv}. We first observe that the gauge invariant local operators $\{{\cal O}(x)\}$ are completely insensitive to the presence of the Stueckelberg field $\xi^a$. In fact, since \eqref{intro17} can be regarded as a particular gauge transformation, it immediately follows that
\begin{equation}
{\cal O}(A(x)) =  {\cal O}(A^h(x)) \;. \label{gi}
\end{equation}
Let  us now proceed by elaborating on the correlator
\begin{equation}
\langle {\cal O}(x) {\cal O}(y) \rangle\Big|_{{S_{\text{GZ}}^{\text{loc}}}}^{\text{Landau}}   = \frac{\int [\mathcal{ D}{\Phi}] \; {\cal O}(x) {\cal O}(y) \; e^{- S_{\text{GZ}}^{\text{loc}}}}{ \int [\mathcal{ D}{\Phi}]   \; e^{- S_{\text{GZ}}^{\text{loc}}  } }  \;, \label{cr}
\end{equation}
where $[\mathcal{D}{\Phi}] $ is again a short-hand notation\footnote{Note that, even though we employ the same notation here and in eq.\eqref{z3}, the integration measures are different.} for integration over all fields entering $S_{\text{GZ}}^{\text{loc}}$, namely the fields already present in eq.\eqref{z3}, $A_{\mu},b,c,\bar c, \varphi,\bar{\varphi},\omega,\bar{\omega}$, as well as the new localizing fields $\xi,\tau,\eta,\bar \eta$. Integration  over $(b, \tau, {\bar \eta}, \eta)$ yields the factor
\begin{equation}
\delta(\partial A) \delta(\partial A^h) \det(\partial D(A^h)) \;. \label{f1}
\end{equation}
Therefore, using the solution $\xi_s$, eq.\eqref{xis}, of the constraint $\partial_\mu A^h_\mu = 0$, we get
\begin{equation}
 \delta(\partial A^h) =  \frac{\delta(\xi-\xi_s)}{| (\partial A^h)'|_{\xi_s}} \;, \label{a11}
\end{equation}
where
\begin{equation}
(\partial A^h)'_{\xi_s}   =  \det( - \partial^2 \delta^{ac} - gf^{abc} (\partial A^b) - gf^{abc} A^b_\mu \partial_\mu + {\cal R}^{ac}(\xi_s) ) \;,   \label{a1}
\end{equation}
with
\begin{equation}
{\cal R}^{ac} = -gf^{acb}(\partial_\mu D^{bd}_\mu \xi_s^ d)-gf^{acb}(D^{bd}_\mu \xi_s^ d)\partial_\mu-\frac{g^2}{2}f^{acb}f^{bde}\left[\partial_\mu(\xi_s^d D^{ef}_\mu \xi_s^f)\right]-\frac{g^2}{2}f^{acb}f^{bde}(\xi_s^d D^{ef}_\mu \xi_s^f)\partial_\mu + \mathcal{O}(\xi^3_s)
\end{equation}
and ${\cal R}(\xi_s)$ collects all remaining infinite power series terms in $\xi_s$. From eq.\eqref{xis} it is clear that $\xi_s$ contains a factor of $\partial A$ for each term in its expansion and, as a consequence, so does ${\cal R}$.\\\\Moreover, taking into account the presence of the delta function $\delta(\partial A)$ stemming from the integration over the field $b$, it follows that (see eqs.\eqref{exA} and \eqref{xis})
\begin{eqnarray}
\xi_s & = & 0 \;, \qquad A^h = A \;,\qquad
(\partial A^h)'_{\xi_s}   =   \det( - \partial^2 \delta^{ac}  - gf^{abc} A^b_\mu \partial_\mu  ) \;.  \label{a3}
\end{eqnarray}
We can therefore remove the modulus $|..|$ in equation \eqref{a11} since, from eq.\eqref{a3}, $(\partial A^h)'_{\xi_s}$ equals precisely the Faddeev-Popov operator of the Landau gauge which turns out to be  positive within $\Omega$. Therefore,
\begin{equation}
\frac{\det(\partial D(A^h))}{ (\partial A^h)'_{\xi_s} } = 1 \;, \label{a4}
\end{equation}
leaving us with a trivial integration over $\xi$ which, due to the delta-function $\delta(\xi)$ of eq.\eqref{a11},  amounts to setting  $\xi=0$ in the remaining expression for the integrand of eq.\eqref{cr}. This establishes the equivalence  \eqref{equiv}. Let us end this section by remarking that  property \eqref{equiv} extends without any difficulty to the Refined Gribov--Zwanziger case.

\section{Universal character of Zwanziger's horizon function for correlation functions of gauge invariant operators}
The BRST invariant reformulation of the Gribov--Zwanziger action enables us to  move to a more general class of gauge fixings. Let us consider in fact the following action
\begin{eqnarray}
{\tilde S} &=&S_{\text{YM}}+S_{\text{GF}}-\int \d^4x\left(\bar{\varphi}^{ac}_{\mu}{{\cal M}^{ab}}(A^h) {\varphi}^{bc}_{\mu}-\bar{\omega}^{ac} {\cal M}^{ab}(A^h)\omega^{bc}_\mu\right) \nonumber \\
\; \; \; \; &-&\gamma^{2}\int \d^4x~gf^{abc}(A^h)^a_{\mu}(\varphi+\bar{\varphi})^{bc}_\mu
+\int \d^4x~\left( \tau^a \partial_\mu (A^h)^a_\mu - \bar{\eta}^a {\cal M}^{ab}(A^h)\eta^b \right) \,,
\label{shat}
\end{eqnarray}
where $S_{\text{GF}}$ stands for an arbitrary covariant, color invariant and renormalizable gauge fixing. To offer an explicit example of what  $S_{\text{GF}}$ might look like, we may consider the following choice
\begin{eqnarray}
S_{\text{GF}} &=&  \int \dd^4x \;s \left( {\bar c}^a (\partial_\mu A^a_\mu - \mu^2 \xi^a + \frac{g}{2} \beta f^{abc} {\bar c}^b c^c) - {i} \frac{\alpha}{2} {\bar c^a} b^a \right)  \nonumber \\
&=&  \int \dd^4x \left( i b^a \partial_\mu A^a_\mu +  \frac{\alpha}{2} b^a b^a
- i \mu^2 b^a \xi^a +ig\beta f^{abc} b^a {\bar c}^b c^c +  \frac{g^2}{4} \beta f^{abc}f^{cmn} {\bar c}^a {\bar c}^b c^m c^n \right) \nonumber  \\
\; \; \; \;&+& \int \dd^4x \left(  {\bar c}^a \partial_\mu D^{ab}_\mu(A)c^b + \mu^2 {\bar c}^a g^{ab}(\xi) c^b  \right)    \;, \label{gf1}
\end{eqnarray}
with  $g^{ab}(\xi)$  is the BRST-transformation of $\xi$, see \cite{Capri:2017bfd}. Evidently, we have $s {\tilde S} = 0$. The expression \eqref{gf1} contains three gauge parameters $\sigma_i= (\alpha, \beta, \mu^2)$, as can be stated in terms of the cohomology of the BRST operator $s$, {\it i.e.}
\begin{equation}
\frac{\partial {\tilde S}}{\partial \sigma_i} = s \Delta_i \;, \qquad i=1,2,3 \;,   \label{trv}
\end{equation}
for some local integrated $\Delta_i$. Equation \eqref{trv} expresses the fact that, unlike the parameters $(\gamma^2, m^2, M^2)$ of the GZ and RGZ actions, the parameters $\sigma_i= (\alpha, \beta, \mu^2)$ are associated to unphysical trivial elements of the cohomology of the BRST operator, see \cite{Piguet:1995er}. \\\\As it is apparent from eq.\eqref{gf1}, both Lorentz covariance and global color invariance are preserved. Furthermore, setting $\sigma_i= (\alpha, \beta, \mu^2)=0$, the Landau gauge is recovered. Besides, when  $(\beta, \mu^2)=0$, expression \eqref{gf1} yields the class of the linear covariant gauges  \cite{Capri:2017bfd} while, for $\beta=0$, gives the class of  $R_\xi$-gauges considered in \cite{Capri:2017npq}. Finally, when $\mu^2=0, \;\beta=\frac{\alpha}{2}$, the Curci-Ferrari non-linear gauges are recovered \cite{Capri:2017abz}.  The all order renormalizability of the action ${\tilde S}$, eq.\eqref{shat}, can be achieved by repeating the specific algebraic analysis already done in the cases of the linear covariant gauges \cite{Capri:2017bfd} and of the $R_\zeta$-gauges \cite{Capri:2017npq}. Of course, more general gauge fixings containing more gauge parameters can be envisaged, without altering the properties we are going to  establish in the following.  \\\\Due to the gauge nature of $\sigma_i= (\alpha, \beta, \mu^2)$, eq.\eqref{trv}, it follows that they will not enter the quantum corrections affecting the parameters $(\gamma^2,m^2, M^2)$, as the latter are linked to nontrivial elements of the cohomology of the BRST operator $s$   \cite{Piguet:1995er}, eqs.\eqref{coh1}. Said otherwise, the anomalous dimensions of $(\gamma^2,m^2, M^2)$ will be independent of $\sigma_i= (\alpha, \beta, \mu^2)$. \\\\Let us now give a look at the correlation functions of the local invariant operators ${\cal O}(x)$ evaluated with the action ${\tilde S}$, namely
\begin{equation}
\langle {\cal O}(x) {\cal O}(y) \rangle\Big|_{{\tilde S}}  = \frac{\int [{\cal D}{\Phi}] \; {\cal O}(x) {\cal O}(y) \; e^{- {\tilde S}}}{ \int [{\cal D}{\Phi}]   \; e^{- {\tilde S}  } }  \;. \label{ooh}
\end{equation}
Since, by definition, the operators ${\cal O}(x)$ are elements of the BRST cohomology
\begin{equation}
s  {\cal O}(x) = 0 \;,  \qquad {\cal O}(x) \neq s {\cal {\hat O}}(x) \;, \label{noo}
\end{equation}
for any local $ {\cal {\hat O}}(x)$, it follows that
\begin{eqnarray}
\frac{\partial \langle {\cal O}(x) {\cal O}(y) \rangle\Big|_{{\tilde S}} }{\partial \sigma_i}  &=&  - \frac{\int [{\cal D}{\Phi}] \;  s\left( {\cal O}(x) {\cal O}(y) \Delta_i \; e^{- {\tilde S}} \right) }{ \int [{\cal D}{\Phi}]   \; e^{- {\tilde S}  } } \nonumber \\
&+&  \left( \frac{\int [{\cal D}{\Phi}] \;   {\cal O}(x) {\cal O}(y) \; e^{- {\tilde S}}  }{ \int [{\cal D}{\Phi}]   \; e^{- {\tilde S}  } } \right) \left( \frac{\int [{\cal D}{\Phi}] \;  s\left(  \Delta_i \; e^{- {\tilde S}} \right) }{ \int [{\cal D}{\Phi}]   \; e^{- {\tilde S}  } } \right) = 0 \;,  \label{oo1}
\end{eqnarray}
where use has been made of eq.\eqref{trv} and of the BRST invariance of ${\tilde S}$. Equation \eqref{oo1}  states that the correlation functions $\langle {\cal O}(x) {\cal O}(y) \rangle\Big|_{{\tilde S}} $ are independent of the gauge parameters $\sigma_i= (\alpha, \beta, \mu^2)$, a feature which directly follows from the BRST invariance. Therefore, without loss of generality, $\langle {\cal O}(x) {\cal O}(y) \rangle\Big|_{{\tilde S}} $  can be evaluated by setting immediately $\sigma_i= (\alpha, \beta, \mu^2)=0$, namely
\begin{equation}
\langle {\cal O}(x) {\cal O}(y) \rangle\Big|_{{\tilde S}}  = \left( {\langle {\cal O}(x) {\cal O}(y) \rangle\Big|_{{\tilde S}}} \right)_{\sigma_i=0} = \langle {\cal O}(x) {\cal O}(y) \rangle\Big|_{{S_{\text{GZ}}^{\text{loc}}}}^{\text{Landau}} = \langle {\cal O}(x) {\cal O}(y) \rangle\Big|_{S_{\text{GZ}}}^{\text{Landau}} \;, \label{r1}
\end{equation}
as it follows from the equivalence \eqref{equiv}  established in the previous section. \\\\Equation \eqref{r1} summarises the main result of the present work, stating that the correlation functions of local BRST invariant operators are independent of the gauge parameters entering the gauge fixing condition. This property holds for a generic class  of gauge condition, provided covariance and global color invariance are  maintained together with the requirement that the Landau gauge is recovered when all gauge parameters are set to zero. This property gives to Zwanziger's horizon function $H(A)$, eq.\eqref{intro8}, a universal character as far as physical correlators are concerned.\\\\ As a motivation for the introduction of the horizon function \eqref{Hh} to remove Gribov copies in the general class of gauges defined by \eqref{gf1} one can employ an argument already advocated in \cite{Pereira:2016fpn}. In the case where the gauge parameters $(\beta,\mu^2)$ are re-expressed as $(\beta,\mu^2)=(\alpha\tilde{\beta},\alpha\tilde{\mu}^2)$, one can write
\begin{equation}
S_{\text{GF}} =  s\int \dd^4x \;  {\bar c}^a\left[ \partial_\mu A^a_\mu -i\frac{\alpha}{2}\left(b^a-2i\tilde{\mu}^2\xi^a + i g \tilde{\beta} f^{abc} {\bar c}^b c^c \right)  \right]\,.
\label{redef1}
\end{equation}
By a suitable redefinition of the $b$ field, namely
\begin{equation}
b'^a = b^a-2i\tilde{\mu}^2\xi^a + i g \tilde{\beta} f^{abc} {\bar c}^b c^c\,,
\label{redef2}
\end{equation}
with trivial Jacobian, expression \eqref{redef2} can be recast as
\begin{equation}
S_{\text{GF}} =  s\int \dd^4x \;  {\bar c}^a\left( \partial_\mu A^a_\mu -i\frac{\alpha}{2}b'^a  \right)\,.
\label{redef3}
\end{equation}
This gauge fixing action is formally equivalent to the linear covariant gauges\footnote{Of course, the action  \eqref{redef3} is not the one of the linear covariant gauges at the dynamical level since the BRST transformation for $b$ is modified due to the  shift \eqref{redef2}, see \cite{Pereira:2016fpn} for a similar observation  in the specific case of the Curci-Ferrari gauge.} and, as such, one can employ the same arguments worked out in \cite{Capri:2015ixa} for the elimination of Gribov copies. In fact, this reasoning does not depend on the particular form of \eqref{gf1}. If one chooses a covariant and color-invariant gauge which can be expressed as
\begin{equation}
S_{\text{GF}} =  s\int \dd^4x \;  {\bar c}^a\left( \partial_\mu A^a_\mu + \alpha \Omega^a  \right)\,,
\label{redef4}
\end{equation}
with $\Omega^a$ a general function of the fields and their derivatives with dimension two and ghost number zero and which,  after a proper rescaling of all gauge parameters,  reduces to the Landau gauge in the limit  $\alpha\to 0$\footnote{Assumed here to exist and to be taken.}, the same argument applies.  What we are effectively doing is restricting the integration of gauge fields to those configurations that --evidently-- obey the chosen gauge condition, supplemented with the constraint that $\mathcal{M}(A^h)>0$, the latter being encoded in the (BRST invariant) Gribov--Zwanziger action and horizon condition. \\\\Let  us end this section with an interesting remark about the role played by the massive gauge parameter $\mu^2$ entering the gauge condition \eqref{gf1}.  As it is clear from its  dimensionful  nature, this parameter provides a BRST invariant regularizing infrared mass to the dimensionless Stueckelberg field $\xi^a$, namely
\begin{equation}
\langle \xi^a(k) \xi^b(-k) \rangle = \frac{\delta^{ab} \alpha}{(k^2 +\mu^2)^2}   \;. \label{xip}
\end{equation}
In the absence of the parameter $\mu^2$, the Stueckelberg field would behave like $\langle \xi^a \xi^a \rangle_k \sim \frac{1}{k^4}$, which might give rise to potential IR spurious divergences in some class of Feynman diagrams. Nevertheless, the possibility of introducing a fully BRST invariant regularizing mass $\mu^2$ through the gauge fixing together with the results \eqref{oo1} and \eqref{r1}  imply that the physical correlation functions are perfectly free from potential IR divergences even when the Stueckelberg field is massless, {\it i.e.} when $\mu^2=0$.  This can also be appreciated from the observation that, when specifying to the Landau gauge, the Stueckelberg field actually has a null propagator and it completely decouples from the theory. Finally, properties  \eqref{oo1},\eqref{r1} extend immediately to the Refined Gribov--Zwanziger  framework.

\section{Conclusion}
In this work we have exploited the recent local BRST invariant reformulation of the Gribov--Zwanziger framework and of its Refined version \cite{Capri:2015ixa,Capri:2016aqq,Capri:2016gut,Capri:2017bfd}.
The existence of an exact nilpotent BRST symmetry has far-reaching consequences, encoded in the powerful language of the cohomology of the corresponding BRST operator $s$, see \cite{Piguet:1995er} and references therein. \\\\The BRST symmetry enables us to attach a clear physical meaning to the Gribov parameter $\gamma^2$ as well as to the parameters $(m^2,M^2)$ entering the Gribov--Zwanziger action and its Refined version, eqs.\eqref{intro10},\eqref{intro11}. Being related to nontrivial elements of the cohomology of the BRST operator $s$, these parameters will not be  affected by the gauge parameters entering the gauge condition to all orders, a fundamental property in order to be seen as physical parameters.\\\\A second important consequence of the BRST symmetry is that of ensuring that the correlation functions of local BRST invariant operators are  independent of the gauge parameters, a feature which has been proven for a very huge class of gauge fixings, as shown  by eqs.\eqref{oo1},\eqref{r1}.  \\\\Equations \eqref{oo1},\eqref{r1} summarise our main result. They grant a universal character to Zwanziger's horizon function  $H(A^h)$, eq.\eqref{Hh}, the BRST invariant version of its Landau gauge limit $H(A)$, eq.\eqref{intro8}. As  already stated, eqs.\eqref{oo1},\eqref{r1} imply that the correlation functions of BRST local invariant operators evaluated with the Yang--Mills action supplemented with Zwanziger's horizon function and quantized in an arbitrary class of covariant, color invariant and renormalizable gauges which reduce to the Landau gauge when all gauge parameters are set to  zero  have a unique, gauge parameters independent result, corresponding to that of the Landau gauge. As a consequence, the restriction to the Gribov region $\Omega$ in the Landau gauge,
eq.\eqref{omega}, acquires a gauge independent meaning in the class of the physical correlators.  \\\\Of course, unlike the BRST invariant correlators, gauge dependent quantities as, for example, the gluon propagator, the three and four gluon vertices, the ghost-gluon vertex and so on, will strongly depend on the specific features and parameters entering the gauge condition. Evidently, for these quantities, the handling of the Gribov issue might be full of highly nontrivial details as one can figure out, for example,  from the case of the linear covariant gauges extensively discussed in  \cite{Capri:2015ixa,Capri:2016aqq,Capri:2016gut,Capri:2017bfd,Capri:2015nzw,Capri:2015pja,Sobreiro:2005vn}.  \\\\Moreover, concerning the gluon propagator, although it is depending on the gauge parameters, it  can be proven that the pole mass of its transverse component is gauge parameter independent to all orders \cite{Capri:2016gut}, thanks to the  so-called Nielsen identities which are again a direct consequence of the BRST symmetry, see \cite{Capri:2016gut} for a detailed derivation of these  identities  within the Gribov--Zwanziger framework. \\\\As a future challenge, we are naturally led to consider a possible extension of the present work to a completely different class of gauge fixings which lack some of the properties which we have required: covariance and/or global color invariance. This is the case of the Coulomb and Maximal Abelian gauges for which the corresponding Zwanziger horizon functions are known, see, for instance,  \cite{Zwanziger:2006sc} and references therein for the Coulomb gauge and  \cite{Capri:2008vk,Capri:2010an,Gongyo:2013rua}  for the Maximal Abelian gauge. The Coulomb gauge lacks covariance while the Maximal Abelian gauge lacks explicit global color invariance. Although expected on physical grounds, the explicit proof of the equivalence of the physical correlation functions among all these gauges would be a remarkable result, due to the highly nontrivial differences that both Coulomb and Maximal Abelian gauges display with respect to the Landau gauge. For example, the lack of covariance of the Coulomb gauge turns the all order proof of the renormalizability of the theory a quite difficult issue due to the appearance of non-local divergences requiring non-local terms \cite{Andrasi:2015cfa}, a feature absent in the Landau and Maximal Abelian gauges. On the other hand, the breaking of the global color invariance of the Maximal Abelian gauge is at the origin of the so-called Abelian dominance  \cite{Ezawa:1982bf,Amemiya:1998jz,Gongyo:2013sha}, according to which the relevant degrees of freedom in the infrared non-perturbative region should be identified with  the Abelian components of the gauge field corresponding to the Cartan subgroup of the  gauge group. In addition, all these gauges lead to
quite different frameworks in order to account for confinement, see \cite{Greensite:2011zz} for a pedagogical general review.  Nevertheless, it might be worth to mention a few encouraging features which might be exploited to figure out  nontrivial checks towards  the possible equivalence among the  physical correlators. As recently done in the case of the Maximal Abelian gauge \cite{Capri:2015pfa,Capri:2017abz}, the gauge invariant field $A^h_\mu$, eq.\eqref{min0}, can be employed to obtain a manifestly BRST invariant reformulation of the Coulomb gauge as well. Also, an interpolating gauge fixing  relating Landau, Coulomb and Maximal Abelian gauges can be found in \cite{Capri:2006bj}, see also \cite{Dudal:2004rx}. This suggests to pursue the idea of trying to built up a kind of generalized interpolating horizon function, a tool which might be helpful to relate the various non-perturbative aspects of all these gauges. Finally, let us mention that, assuming the hypothesis of the Abelian dominance \cite{Ezawa:1982bf}, a first evidence that the spectrum of the lightest glueballs in the Maximal Abelian gauge  is in agreement with that already obtained in
the Landau gauge \cite{Dudal:2010cd,Dudal:2013wja} has been outlined in \cite{Capri:2011ki}. This result can be interpreted in favour of the aforementioned equivalence among the physical correlation functions. Any progress in this direction will be reported soon.

 \section*{Acknowledgements.}
The Conselho Nacional de Desenvolvimento Cient\'{i}fico e
Tecnol\'{o}gico (CNPq-Brazil), the Faperj, Funda{\c{c}}{\~{a}}o de
Amparo {\`{a}} Pesquisa do Estado do Rio de Janeiro, the SR2-UERJ
and the Coordena{\c{c}}{\~{a}}o de Aperfei{\c{c}}oamento de
Pessoal de N{\'\i}vel Superior (CAPES) are gratefully acknowledged
for financial support. A.D.P. acknowledges funding by the DFG, Grant Ei/1037-1.

\end{document}